# Model-based ROC (mROC) curve: examining the effect of case-mix and model calibration on the ROC plot

Mohsen Sadatsafavi[1,2], Paramita Saha-Chaudhuri[3], John Petkau[4]

1. Faculty of Pharmaceutical Sciences, The University of British Columbia
2. Faculty of Medicine, The University of British Columbia
3. Department of Statistics, University of Vermont
4. Department of Statistics, The University of British Columbia

**Corresponding Author:**

Mohsen Sadatsafavi
Associate Professor, Faculty of Pharmaceutical Sciences
The University of British Columbia
http://resp.core.ubc.ca/team/msafavi
msafavi@mail.ubc.ca

**Word count:** 5,223

**Abstract word count:** 292

**Acknowledgement**
We would like to thank Drs. David van Klaveren (Erasmus University and Tufts Medical Center) and Abdollah Safari (University of British Columbia) for their insightful comments on earlier drafts.

**Declaration of Conflicting Interests**

The Authors declare that there is no conflict of interest.

**Abstract**

**Background:** The performance of risk prediction models is often characterized in terms of discrimination and calibration. The Receiver Operating Characteristic (ROC) curve is widely used for evaluating model discrimination. However, when comparing ROC curves across different samples, the effect of case-mix makes the interpretation of discrepancies difficult. Further, compared to model discrimination, evaluating model calibration has not received the same level of attention. Current methods for examining model calibration require specification of smoothing or grouping factors.

**Methods:** We introduce the “model-based” ROC curve (mROC) to assess the external calibration of a prediction model. mROC curve is the ROC curve that should be observed if the prediction model is calibrated in the external population. Unlike the ROC curve, the mROC curve is affected by even monotonic transformations of predicted risks, and thus is sensitive to model calibration. We propose a novel test statistic for calibration that, unlike current methods, does not require any subjective specification of smoothing or grouping factors.

**Results:** Through a stylized example, we demonstrate how mROC separates the effect of case-mix and model miscalibration when comparing a prediction model’s ROC curves from different samples. We present the results of simulation studies that confirm the properties of our new calibration test. A case study on predicting the risk of acute exacerbations of Chronic Obstructive Pulmonary Disease puts the developments in a practical context. R code for the implementation of this method is provided.

**Conclusion:** mROC can easily be constructed and used to interpret the effect of case-mix and calibration on the ROC plot. Given the popularity of ROC curves among applied investigators, this framework can further promote assessment of model calibration.

## Background

Risk prediction models that objectively quantify the probability or risk of clinically important events based on observable characteristics are critical tools for efficient patient care. A risk prediction model is typically constructed in a development (or training) sample, but before it is adopted for use in a target population, its performance needs to be assessed in an independent (external) validation sample drawn from that population. In examining the appropriateness of a risk model, two fundamental aspects are discrimination and calibration. The former refers to the capacity of the model to properly stratify individuals with different risk profiles, and the latter refers to the degree to which predicted risks are close to the true risks(1).

The Receiver Operating Characteristic (ROC) curve and the area under the ROC curve (AUC, or the c-statistic) are classical examples of tools for assessing model discrimination(2). When evaluating a risk prediction model in a sample, the discriminatory performance of the model can be affected by both the distribution of predictor variables (case-mix) as well as the validity of the model in that sample(3). Consequently, when comparing the performance of a model between development and validation samples, differences in the case-mix between the two samples can make comparisons difficult. One area of interest in the present work is to separate these two sources of discrepancy. An early advance in this area was made by Vergouwe et. al. who proposed benchmarks based on simulating responses from predicted risks and fitting the model in the validation sample(4). More recent work has largely focused on the AUC, an overall summary measure of the ROC curve(3,5,6).

Compared to model discrimination, examining model calibration has not received the same level of attention(7,8). Model calibration is often neglected in the evaluation of the overall performance of risk prediction models, so much so that it is referred to as “the Achilles’ heel of predictive analytics”(9). In the context of a logistic model for binary responses, Van Calster et al(10) proposed a hierarchy of definitions for model calibration. In particular, a model is ‘moderately calibrated’ if the average observed risk among all subjects with a given predicted risk is equal to the predicted risk. Moderate calibration is contrasted with ‘weak’ calibration (when a linear calibration plot has an intercept of zero and slope of one), as well as with ‘strong’ calibration (when the predicted and observed risks are equal for all covariate patterns – an unrealistic condition in practical situations)(10). Moderate calibration is typically assessed using the calibration plot, which shows the average value of the observed risk as a function of the predicted risk after grouping or smoothing the response values.

In this work we propose model-based ROC (mROC) analysis as an objective method of assessing model calibration. The mROC enables investigators to disentangle the effect of case-mix and model validity on the shape of the ROC curve. Importantly, we show that the mROC connects ROC analysis, a classical means of evaluating model discrimination, to model calibration. We use this connection to propose a novel test for the assessment of model calibration that does not require specification of smoothing or grouping factors.

**Notation and context**

Our main interest is in the ‘external validation’ context where a previously developed risk prediction model for a binary outcome is applied to a new independent (external) sample to examine its performance in that sample’s target population. The risk prediction model is given by the deterministic function $\pi^*(\mathbf{X})$, mapping an individual’s covariate vector $\mathbf{X}$ to $\pi^*$, the probability of observing the binary outcome (response) of interest (e.g., whether a patient with asthma will experience a flare-up in the next six months). Let $Y$ be the binary outcome of interest, with $Y = 1$ indicating the presence of the disease or the occurrence of the event, and 0 otherwise. In what follows, unless otherwise specified, by “calibration” we refer to moderate calibration, i.e., $P(Y = 1|\pi^*(\mathbf{X}) = p) = p$. Applying this model to the external sample consisting of a random sample of $n$ individuals, we obtain $\boldsymbol{\pi}^* = (\pi_1^*, \dots, \pi_n^*)$, the vector of predicted risks. In the external sample, we also observe the corresponding vector $\mathbf{Y} = (Y_1, \dots, Y_n)$ of response values.

**The empirical ROC curve**

Two fundamental probability distributions underlie the ROC curve: the distribution of predicted risks among individuals who experience the event (positive individuals, or cases), and among individuals who do not experience the event (negative individuals, or controls). Let $F_1$ and $F_0$ represent the corresponding cumulative distribution functions (CDFs) of the predicted risk:

$$F_1(t) = P(\pi^* \leq t | Y = 1),$$

$$F_0(t) = P(\pi^* \leq t | Y = 0).$$

The true positive (TP) and false positive (FP) probabilities are closely linked with the distribution of risk among the positive and negative individuals, respectively: $TP(t) \equiv P(\pi^* > t | Y = 1) = 1 - F_1(t)$, and $FP(t) \equiv P(\pi^* > t | Y = 0) = 1 - F_0(t)$. The population ROC curve induced by the risk prediction model $\pi^*$ can be expressed as

$$ROC(t) = 1 - F_1(F_0^{-1}(1 - t)),$$

where $0 \leq t \leq 1$ is the false positive probability(11).

With the external dataset, consistent estimators for $F_1$ and $F_0$ can be obtained by averaging the indicators $I(\pi_i^* \leq t)$ for each of the positive and negative groups:

$$F_{1n}(t) = \frac{\sum_{i=1}^{n}\{I(\pi_i^* \leq t).Y_i\}}{\sum_{i=1}^{n} Y_i},$$

and

$$F_{0n}(t) = \frac{\sum_{i=1}^{n}\{I(\pi_i^* \leq t).(1 - Y_i)\}}{n - \sum_{i=1}^{n} Y_i}.$$

$F_{1n}(t)$ and $F_{0n}(t)$ are used to generate $ROC_n(t)$, the empirical ROC, as a consistent estimator of the population ROC curve(12,13).

**The model-based ROC (mROC) curve**

The $i^{\text{th}}$ subject in the external sample is a random draw from the set of all individuals in the target population whose predicted risk is $\pi_i^*$. Hence, under the assumption that the model is calibrated, we have $P(Y_i = 1) = P(Y = 1 | \pi^* = \pi_i^*) = \pi_i^*$; that is, the vector of observed response values is a random draw of independent Bernoulli trials from the vector of predicted risks. Hence, in addition to the ROC curve based on the observed responses, one can also construct a ROC curve based on the potential random responses generated from the Bernoulli distribution with probabilities equal to the predicted risk.

Let $Y^*$ be a random realization of this potential response from the predicted risk of a randomly selected individual. The ROC-related CDFs based on $Y^*$ are:

$$\bar{F}_1(t) = P(\pi^* \leq t | Y^* = 1),$$

and

$$\bar{F}_0(t) = P(\pi^* \leq t | Y^* = 0).$$

The application of Bayes' rule leads to the following estimators in the external sample:

$$\bar{F}_{1n}(t) = \frac{\sum_{i=1}^{n} I(\pi_i^* \leq t\ ).\pi_i^*}{\sum_{i=1}^{n} \pi_i^*},$$

and

$$\bar{F}_{0n}(t) = \frac{\sum_{i=1}^{n} I(\pi_i^* \leq t\ ).(1-\pi_i^*)}{n - \sum_{i=1}^{n} \pi_i^*}.$$

Hence, one can generate a 'model-based' ROC or $mROC_n(t)$, independently of the observed outcomes in the external sample, based on the CDFs $\bar{F}_{1n}$ and $\bar{F}_{0n}$ obtained by averaging the indicator functions $I(\pi_i^* \leq t\ )$ with weights of $\pi_i^*/\sum \pi_i^*$ and $(1-\pi_i^*)/\sum(1-\pi_i^*)$ for the *i*$^{th}$ individual in the sample. Therefore, mROC is the "expected" ROC for the external sample based on predictions obtained from the risk prediction model. This is extension of the definition of model-based c-statistic proposed by van Klaveren et al to the entire ROC curve(3). As demonstrated in ***Supplementary Material - Section 1***, the area under the mROC curve is equal to the model-based c-statistic(3).

**The connection between mROC curve, case-mix, and model calibration**

The limiting forms of the estimated CDFs $F_{1n}$, $F_{0n}$, $\bar{F}_{1n}$ and $\bar{F}_{0n}$ are derived in ***Supplementary Material - Section 2***. An important consequence is that, provided that the model is calibrated in the external sample, $ROC_n(t)$ and $mROC_n(t)$ converge to the same value at each point $t$, as $n$, the sample size in the external sample, approaches infinity. That is, moderate calibration is a sufficient condition for convergence of the empirical ROC and mROC curves. A stylized example demonstrating this is provided in ***Supplementary Material – Section 3***.

Unlike in the expression of $F_{1n}$ and $F_{0n}$, the observed outcomes in the sample do not appear in the expression of $\bar{F}_{1n}$ and $\bar{F}_{0n}$. The behavior of these CDFs depends on the predicted risks, rather than the observed outcomes in the sample. Therefore, the mROC curve depicts the case-mix-adjusted ROC curve: the ROC curve that would be expected to be observed in the sample, if the model is calibrated in this sample. This motivates our proposal for using mROC to gain insight into the effect of case-mix and model calibration when examining the external validity of a model.

Consider the mROC and empirical ROC curves in the validation sample when examining the external validity of a model. The former carries the association between the predictors and outcome from the development sample through the prediction model, whereas the latter captures such association in the validation sample. However, both are based on the case-mix in the validation sample. Because of the shared case-mix, discrepancies between these curves point toward model miscalibration in the validation sample. This can be demonstrated using a stylized example: We consider a single predictor $X$, which has a standard normal distribution in the development population. Using a sample from the development population, we construct a risk prediction model as $P(Y = 1) = 1/(1 + exp(-X))$, which happens to be the correctly specified model (and thus is calibrated) in this population. This model has an AUC of 0.740 in the development population. Now consider four hypothetical external validation scenarios. In the first scenario (***Figure 1***, panel A), the distribution of $X$ and its association with the outcome are the same in the validation population as in the development population. As such, the external (empirical) ROC and mROC curves agree (and will also resemble the development ROC curve). In the second scenario (***Figure 1***, panel B), the predictor is under-dispersed in the validation population (s.d.=0.5), while the association is still the same, thus the model is calibrated. Given the lower variance of risks, the model has lower discriminatory power in this population (AUC=0.641). Both the empirical ROC and mROC curves move closer to the diagonal line, but they closely match each other. Next, consider a validation population that has the same distribution of $X$ as the development population, but with a weaker predictor-outcome association ($P(Y=1)=1/(1+exp(-X/2))$ – thus the model is 'optimistic' and not calibrated). This again causes the empirical ROC curve to be closer to the diagonal line (***Figure 1***, panel C, AUC=0.641). Here, however, the mROC curve remains unchanged from the first scenario. This pattern indicates that the change in the discriminatory performance of the model between the development and validation samples is due to model miscalibration in the validation sample. Finally, consider a validation population in which the predictor is under-dispersed and the association is weaker (***Figure 1***, panel D). Both factors contribute to the empirical ROC curve being closer to the diagonal line (AUC=0.584). Here, due to the difference in the case-mix, the mROC curve also gets closer to the diagonal line, but due to the mis-calibrated model in the validation sample, it is not aligned with the empirical ROC curve. This demonstration implies that difference in case-mix between the development and validation samples does not lead to the discrepancy between the mROC curve and the empirical ROC curve; however, miscalibration of the prediction model in the external sample can lead to discrepancy.

<<Figure 1>>

**mROC as the basis of a novel statistical test for model calibration**

While moderate calibration is a sufficient condition for the convergence at all points of the empirical ROC and mROC curves, moderate calibration on its own is not a necessary condition for such convergence. To progress, in ***Supplementary Material - Section 4*** we show that at the population level, the equivalence of ROC and mROC curves guarantees moderate calibration if an additional condition is imposed. This condition is mean calibration, i.e., $E(\pi^*)=E(Y)$, a condition whose assessment is an integral part of external validation of a risk prediction model(14).

To examine such population-level quantities in a sample, we propose a statistical inference procedure. We define the null hypothesis ($H_0$) as the model being calibrated: $P(Y = 1|\pi^* = p) = p$. Given the results in ***Supplementary Material - Section 4***, $H_0$ can be seen as a combination of two null hypotheses, one on the equivalence of the expected values of predicted and observed risks ($H_{0A}$), and the other on the equivalence of the mROC and ROC curves ($H_{0B}$):

$$H_0: \begin{cases} H_{0A} & E(\pi^*) = E(Y) & \text{mean calibration} \\ H_{0B} & \forall t \; mROC(t) = ROC(t) & \text{mROC and ROC equality.} \end{cases}$$

These hypotheses jointly provide the necessary and sufficient conditions for the risk prediction model to be calibrated.

For $H_{0A}$, consider $A = |E(Y) - E(\pi^*)|$. This population quantity achieves its minimum value of 0 if $H_{0A}$ is true. Our proposed test statistic is the sample estimator of this quantity, the absolute average distance between the observed and predicted risks in the sample:

$$A_n = \frac{1}{n} . \left| \sum_{i=1}^{n} (Y_i - \pi_i^*) \right| \qquad \text{(mean calibration statistic).}$$

For $H_{0B}$, consider the population quantity $B = \int_0^1 |ROC(t) - mROC(t)|. dt$, which achieves its minimum value of 0 when the ROC and mROC curves are equal at all points. Our proposed test statistic is a sample estimator for this quantity, the integrated absolute difference between the empirical ROC and mROC curves in the sample:

$$B_n = \int_0^1 |ROC_n(t) - mROC_n(t)|. dt \qquad \text{(ROC equality statistic).}$$

Given that both $ROC_n$ and $mROC_n$ are step functions, the above integral is the sum of rectangular areas and can be evaluated exactly.

The null distributions of both $A_n$ and $B_n$ can be approximated numerically through straightforward Monte Carlo simulations. Through simulating vectors of response values from the vector of predicted probabilities, one can generate many simulated ROC curves and use them to construct empirical distribution functions under $H_0$ for $A_n$ and $B_n$. These empirical distributions can then be used to generate approximate one-tailed p-values for these two statistics as:

$$p_{A_n} = 1 - eCDF_{A_n}(A_n),$$

where $eCDF_{A_n}$ is the empirical CDF of the mean calibration statistic under $H_0$, and

$$p_{B_n} = 1 - eCDF_{B_n}(B_n),$$

where $eCDF_{B_n}$ is the empirical CDF of the ROC equality statistic under $H_0$.

Individually, the two statistics provide insight about the performance of the model. However, it is more desirable to obtain a single overall p-value for $H_0$. If these tests were independent, one could use Fisher's method(15) to obtain a unified p-value, as under $H_0$, $p_{A_n}$ and $p_{B_n}$ have standard uniform distributions; thus the statistic

$$U_n = -2.\left[\log(p_{A_n}) + \log(p_{B_n})\right]$$

would have a chi-square distribution with 4 degrees of freedom(16). However, as the two statistics are generated from the same data, they are dependent. An adaptation of Fisher's method for dependent P values (based on matching the moments of the test statistic to that of a chi-square distribution) can be used (17). The steps for generating a unified p-value are outlined in the algorithm provided in ***Supplementary Material – Section 5***.

**Simulation Studies**

We performed simulation studies to evaluate the finite-sample properties of the proposed test and its performance against the conventional Hosmer-Lemeshow and Likelihood Ratio tests. We modeled a single predictor $X$ with a standard normal distribution, and the true risk as $p = 1/(1 + exp(-X))$. We evaluated the performance of the tests in a simulated independent sample of $n$ observations when the predicted risks suffer from various degrees of miscalibration. Two sets of simulations were performed. In the first set, we assumed the risk model generated potentially miscalibrated predictions in the form of $logit(\pi^*) = a + b.logit(p) = a + b.X$. Given the linear association on the logit scale between the predicted and actual risks, the weak and moderate calibrations are equal in these scenarios, and therefore the Likelihood Ratio test (simultaneously testing whether $a = 0$ and $b = 1$) has the maximum theoretical power in detecting miscalibration. As such, this simple setup provides an opportunity to judge the performance of the unified test against a gold standard.

In the second set, the true risk model remained the same as above, and we modeled non-linear miscalibrations as $logit(\pi^*) = a + b.sign(X).|X|^{1/b}$. Here, $a$ affects the mean calibration, while the term involving $b$ is an odd function that flexibly changes the calibration slope but preserves the expected value of the predicted risks. We simulated response values and predicted risks with values $a = \left\{0, \frac{1}{4}, \frac{1}{2}\right\}$ and $b = \left\{\frac{1}{3}, \frac{2}{3}, 1, \frac{4}{3}, \frac{5}{3}\right\}$, and three different sample sizes: $n = \{100, 250, 1000\}$, in a fully factorial design (45 simulation scenarios). ***Figure 2*** presents the population-level calibration plots for each of the unique combination of $a$ and $b$.

<<Figure 2>>

We calculated the power of the mean calibration test, the ROC equality test, the unified test, the Hosmer-Lemeshow test (based on 10 groups), and the Likelihood Ratio test in detecting miscalibration at the 0.05 significance level. Following recommendations on objectively deciding on the number of simulations(18), we obtained the results through 2,500 Monte Carlo iterations such that the maximum S.E around the probability of rejecting $H_0$ ($\sqrt{p(1-p)/nSim}$) would be 0.01. Within each iteration, p-values were calculated from $eCDF_{A_n}$ and $eCDF_{B_n}$that were in turn based on 10^5 simulations. We used R for this analysis(19), with the implementation of the simulation-based estimation of $eCDF_{A_n}$ and $eCDF_{B_n}$ in C for computational efficiency.

Results of the first set of simulations are provided in ***Supplementary Material – Section 6***. The power of the unified test was very close to that of the Likelihood Ratio test across all scenarios examined. ***Figure 3*** provides the empirical ROC and mROC curves for the second set of simulations. As all the mappings from $p$ to $\pi^*$ in these simulations are monotonic, the ROC curve remains the same in all panels (with an AUC of 0.740). However, the mROC is generally affected by miscalibration.

<<Figure 3>>

The performance of all tests are summarized in ***Figure 4.*** The middle panel on the top row, where $a = 0$ and $b = 1$, pertains to the only scenario where $H_0$ is true. All tests appropriately rejected the null hypothesis around the nominal type I error rate of 0.05. Focusing on the first row, given $a = 0$, $E(\pi^*) = E(Y) = 0.5$ under these transformations; thus $A_n$ (mean calibration, the white bars) fails. On the other hand, in the third column, where $b = 1$, thus the predicted odds are proportional to the true odds, $B_n$ (mROC/ROC equality, gray bars) fails, as the mROC and ROC curves are very close to each other under these scenarios (***Figure 3***). However, in all scenarios, the unified test appropriately detected miscalibration. In general, the power of the unified test was either equal to or higher than the power of the Hosmer-Lemeshow and Likelihood Ratio tests. The latter, being a test of weak calibration, can have low power when the miscalibration is S-shape such that the calibration slope remains unchanged (e.g., in the top left panel when a=0 and b=1/3, with 22% power with a sample size of 1,000, compared to >99% power for the unified and Hosmer-Lemeshow tests).

<<Figure 4>>

**Application**

Chronic Obstructive Pulmonary Disease (COPD) is a common chronic disease of the airways. Periods of intensified disease activity, referred to as exacerbations, are an important feature of the disease. Individuals vary widely in their tendency to exacerbate(20). Predicting who is likely to experience an exacerbation, especially a severe one that will require hospital admission, will provide opportunities for preventive interventions(21).

We used data from the MACRO(22) and STATCOPE(23), two clinical trials in COPD patients with exacerbations as the primary outcome, to, respectively, develop and validate a risk prediction model for the occurrence of COPD exacerbations in the first six months of follow-up. Baseline characteristics of both samples are provided in ***Table 1***.

<<Table 1>>

Of note, these data have previously been used for a more sophisticated prediction model(24). Here, we focus on a simpler approach as the nuances of model development are beyond the scope of this work. We used a logistic regression model based on the data from the MARCO trial that included the predictors as listed in ***Table 1*** based on an *a priori* list of covariates generated from prior knowledge of possible association with the outcome. We considered two outcomes: all exacerbation and severe exacerbations, and developed two distinct models. The regression coefficients for both models are provided in ***Table 2***. The study was approved by the University of British Columbia and Providence Health Research Ethics Board (H11–00786).

<<Table 2>>

***Figure 5*** provides the empirical ROC curve from the development sample (MARCO) as well as the empirical ROC and mROC curves from the validation sample (STATCOPE) and the calibration plot for both outcomes. For all exacerbations, the mROC curve was very close to the development ROC curve but not to the external (empirical) ROC curve. This indicates that the reduction in the discriminatory performance of the model in the validation sample is due to miscalibration. Indeed, both components of `the proposed test indicated departure from calibration: the mean calibration test had $p_{A_n} < 0.001$ (a two-tailed t-test also had $p < 0.001$); this was also the case for the equivalence of the mROC and empirical ROC curves in the validation sample ($p_{B_n} < 0.001$). The unified test also rejected the hypothesis that the model is calibrated ($p_{U_n} < 0.001$). The Hosmer-Lemeshow test also produced a p-value of <0.001.The calibration plot suggested miscalibration in the external sample, with a general overestimation of risk.

<<Figure 5>>

The model for severe exacerbations had higher discriminatory power. All three ROC curves were generally aligned with each other. The mean calibration test produced $p_{A_n} = 0.070$ (a two-tailed t-test led to $p = 0.061$), while the ROC equality test resulted in $p_{B_n} = 0.74$. The unified test for model calibration did not indicate evidence against moderate calibration ($p_{U_n} = 0.20$). The Hosmer-Lemeshow test resulted in a p-value of 0.16. The calibration plot suggested generally good agreement between the predicted and observed risks for all but the highest decile of predicted risk (***Figure 5***).

**Discussion**

Our contribution in this manuscript was the introduction of the model-based ROC (mROC) curve, the ROC curve that should be expected if the model is at least moderately calibrated in an external validation sample. We showed moderate calibration is a sufficient condition for the convergence of empirical ROC and mROC curves. We extended these results by proving that together, mean calibration and the equivalence of mROC and ROC curves in the population, are sufficient conditions for the model to be moderately calibrated in the external sample. To test for such equivalences within a sample, we suggested a simulation-based test. To the best of our knowledge, this is the first time that the ROC plot, a classical means of communicating model discrimination, has been connected to model calibration. We have implemented the proposed methodology in an R package, which is available from https://github.com/Shoodood/mROC.

Previous investigators have suggested using case-mix-corrected performance metrics in judging the external validation of a model. Vergouwe et al proposed a general approach for calculating different model-based metrics by simulating responses from predicted risks in the validation sample, and comparing the resulting metrics with the empirical ones based on the observed responses in the validation sample(4). Van Klaveren et al focused on one such metric, the c-statistic, and developed closed-form estimators, that would quantify the expected change in a model's discriminative ability due to case-mix heterogeneity(3). This methodology extends such work to the entire ROC curve, and in doing so it establishes a connection between mROC/ROC equality and model calibration that enables formal statistical inference on moderate calibration. The test that is classically associated with calibration plots is the Hosmer-Lemeshow test, which is criticized due to its sensitivity to the grouping of the data and lack of information about direction of miscalibration(25). Our proposed test is free from arbitrary grouping of the data or the choice of smoothing factors. Our simulations empirically verified the postulated properties of this novel test. Given the shortcomings of the Hosmer-Lemeshow test, alternative inferential techniques for evaluating model calibration have been proposed. Allison reviewed the measures of fit of logistic regression models and categorized them as indices of predictive power (like $R^2$) and goodness of fit(26). In their comprehensive review of goodness-of-fit tests for logistic models(27), Hosmer et al defined goodness-of-fit as the adequacy of a model on three fronts: the link function, the probability distribution, and the linear predictor. But, this is a distinctly different pursuit than examining moderate calibration. Consequently, none of the tests examined by Allison and Hosmer et al can be considered a test for moderate calibration. Our proposed test statistic seems to be the first alternative to the Hosmer-Lemeshow test that strictly examines moderate calibration.

These developments can be used in practice in different ways. Steyerberg and Vergouwe have proposed an “ABCD” approach for external validation of a model (A: mean calibration, B: calibration slope, C: c-statistic, and D: Decision Curve analysis)(14). The B step in this approach can be replaced with the mROC’s B statistic, which, together with the A step (which is the same as the A step in the unified test), will test moderate calibration, the most desired form of calibration, as opposed to weak calibration tested via calibration slope(10). Further, if the research involves simultaneous development and external validation of a model, drawing the empirical ROC and mROC curves will provide visual information on the role of case-mix and model miscalibration on potential differences in the discriminatory performance of the model between the two samples (as demonstrated in our case study). Incompatibility between mROC and empirical ROC in the validation sample will rule out moderate calibration. Conversely, while agreement between the two curves does not rule in moderate calibration *per se*, it does so provided that mean calibration (calibration in the large) is achieved. This visual interpretation can be augmented with formal hypothesis testing using the proposed unified statistic. Such comparisons can also be made for subgroups within the sample, albeit multiple hypothesis testing should be controlled for in such circumstances. Even when the investigators are not planning to produce ROC curves, the proposed test for moderate calibration can be reported independently. This can complement the scalar metrics that measure the degree of miscalibration but are not based on formal hypothesis testing, such as Harrell's Emax(28), Austin and Steyerberg's Integrated Calibration Index(29), and Van Hoorde et al's Estimated Calibration Index(30).

There are several ways the proposed methodology can be extended. The ROC curve has been extended to categorical data(31), as well as to time-to-event data(32,33); similar developments can also be pursued for the mROC methodology. Development of inferential methods that would not require Monte Carlo simulations can also be of potential value. As the ROC curve can be interpreted as a CDF(11), non-parametric statistics based on the distance between CDFs can conceivably be developed to test the equivalence of mROC and ROC curves. However, the calculation of the simulation-based p-value for the ROC equality test is computationally efficient (except for very large datasets). Thus, Monte Carlo error can be made smaller than the error generated from applying asymptotic methods to a finite sample. Further, while we have shown that mROC/ROC compatibility *per se* does not guarantee model calibration, our simulations suggest that such compatibility occurs when predicted and calibrated risks are proportional on the odds scale. As such, mROC/ROC compatibility might mean one should only adjust the intercept term in a logistic regression to achieve moderate calibration. In this sense, our proposed approach has some similarities with the step-wise inferential approach proposed by Vergouwe et al(34) for examining which aspect of a risk prediction model (mean calibration, calibration slope, or individual regression coefficients) need to be updated to improve the performance of the model in a new sample. However, our simulations were proof-of-concept, and this observation should be further corroborated by theoretical developments or more extensive simulations.

One of the promises of Precision Medicine is to empower patients for making informed decisions based on their specific risk of outcomes(35). Basing medical decisions on mis-calibrated predictions can be harmful. Our contribution is the development of mROC analysis, a simple method for separating the effect of case-mix and model miscalibration on the ROC curve, and for inference on model calibration. Recent arguments and counterarguments indicate that the methodological research community is divided in its opinion on the utility of ROC curves in the assessment of risk prediction models (36,37). Given the popularity of ROC curves among applied investigators, these developments can result in more attention to model calibration as an often-neglected but crucial aspect in the development of risk prediction models.

| ***Table 1:*** Baseline characteristics and outcomes for MACRO and STATCOPE samples | | |
|---|---|---|
| **Sample characteristics** | **Development sample (MACRO)** | **Validation sample (STATCOPE)** |
| Sample size | 1,074 | 832 |
| Number (%) with at least one exacerbation during the first six months of follow-up | | |
| All exacerbations | 691 (64.3%) | 454 (54.5%) |
| Severe exacerbations | 141 (13.1%) | 73 (8.8%) |
| Female sex; % | 59.2 | 56.6 |
| Age; mean (IQR) | 65.2 (13.0) | 62.4 (13.0) |
| Previous history of oxygen therapy; % | 59.3 | 48.4 |
| Previous history of hospitalization; % | 50.0 | 31.1 |
| SGRQ; mean (IQR) | 50.1 (22.4) | 49.6 (24.4) |
| $FEV_1$ (liters); mean (IQR) | 1.11 (0.70) | 1.19 (0.81) |
| Current smoker; % | 21.7 | 29.7 |
| Current LABA user, % | 74.4 | 42.6 |
| Current LAMA user, % | 63.5 | 66.1 |

**IQR**: Inter-quartile range; **SGRQ**: St. George Respiratory Questionnaire; **$FEV_1$**: Forced expiratory volume at one second; **LABA**: long-acting beta agonists; **LAMA**: Long-acting anti-muscarinic agents.

| ***Table 2:*** Regression coefficients for the risk prediction models (based on the MACRO sample) for all and severe exacerbations | | |
|---|---|---|
| **Log-odds ratio*** | **All exacerbations**<br><br>**Estimate (SE)** | **Severe exacerbations**<br><br>**Estimate (SE)** |
| Intercept | 0.787 (0.707) | -3.840 (1.018) |
| Female sex | -0.482 (0.145) | 0.209 (0.201) |
| Age (/10)# | -0.094 (0.084) | -0.016 (0.119) |
| Previous history of oxygen therapy | 0.275 (0.147) | 0.297 (0.217) |
| Previous history of hospitalization | 0.490 (0.135) | 0.925 (0.200) |
| SGRQ# | 0.098 (0.043) | 0.219 (0.063) |
| $FEV_1$ (liters)# | -0.158 (0.146) | -0.251 (0.219) |
| Current smoker | -0.168 (0.176) | -0.017 (0.242) |
| Current LABA user | 0.157 (0.155) | 0.466 (0.247) |
| Current LAMA user | 0.354 (0.142) | 0.083 (0.206) |

*We included a coefficient for randomized treatment (azithromycin) but it was set to 0 for prediction (as the model is applicable to those who are not on preventive therapy, and none of the individuals in the validation sample was on such a therapy).

#log-odds ratios are for one unit increase for continuous variables

**SE**: standard error; **SGRQ**: St. George Respiratory Questionnaire; **$FEV_1$**: Forced expiratory volume at one second; **LABA**: long-acting beta agonists; **LAMA**: Long-acting anti-muscarinic agents.

**Figure legends**

***Figure 1:*** Empirical ROC (black) and mROC (red) curves for the stylized example.

***Figure 2:*** Relationship between predicted (X axis) and true (Y axis) risks for the simulation scenarios.

***Figure 3:*** ROC (black) and mROC (red) curves for the simulation scenarios. The panels positionally correspond to the calibration plots and simulation parameters presented in Figure 2.

***Figure 4:*** Probability of rejecting the null hypothesis for the mean calibration (pink bars), ROC equality (orange bars), and unified (purple bars) test statistics. The panels positionally correspond to the calibration plots and simulation parameters presented in Figure 2.

***Figure 5:*** The empirical ROC curves from the MACRO development (blue) and STATCOPE validation (black) samples, the mROC curve from the STATCOPE validation sample (red) (left panels) and the calibration plot (right panels).

***Figure 1:*** Empirical ROC (black) and mROC (red) curves for the stylized example.

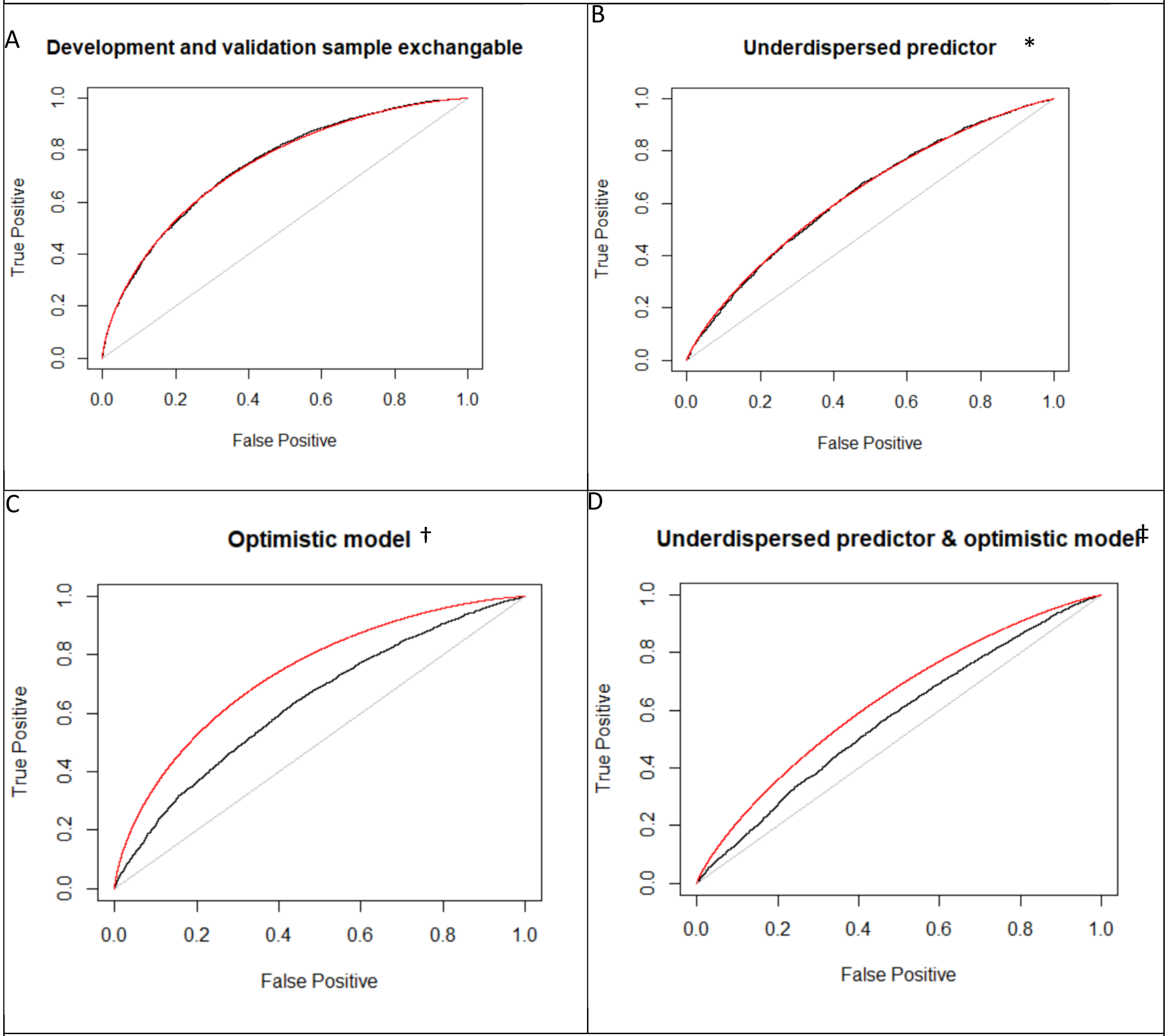

* Distribution of the single predictor in the validation population: $X \sim Normal(\mu = 0, \sigma = 0.5)$

† The association model in the validation population: $P(Y = 1) = 1/(1 + exp(-X/2))$

‡ Predictor distribution same as in panel B, and association model same as in panel C

***Figure 2:*** Relationship between predicted (X axis) and true (Y axis) risks for the simulation scenarios.

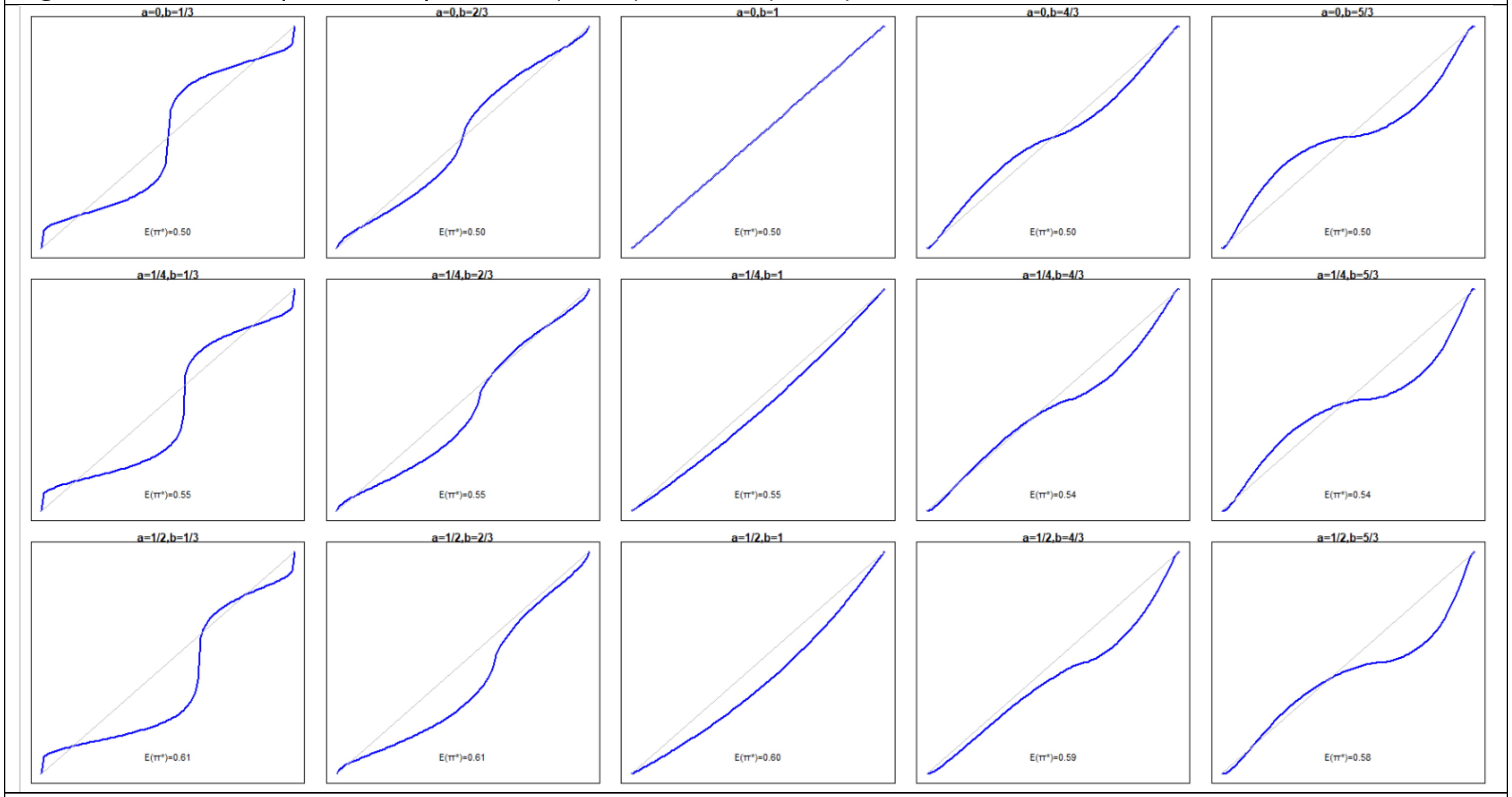

***Figure 3:*** ROC (black) and mROC (red) curves for the simulation scenarios. The panels positionally correspond to the calibration plots and simulation parameters presented in Figure 2.

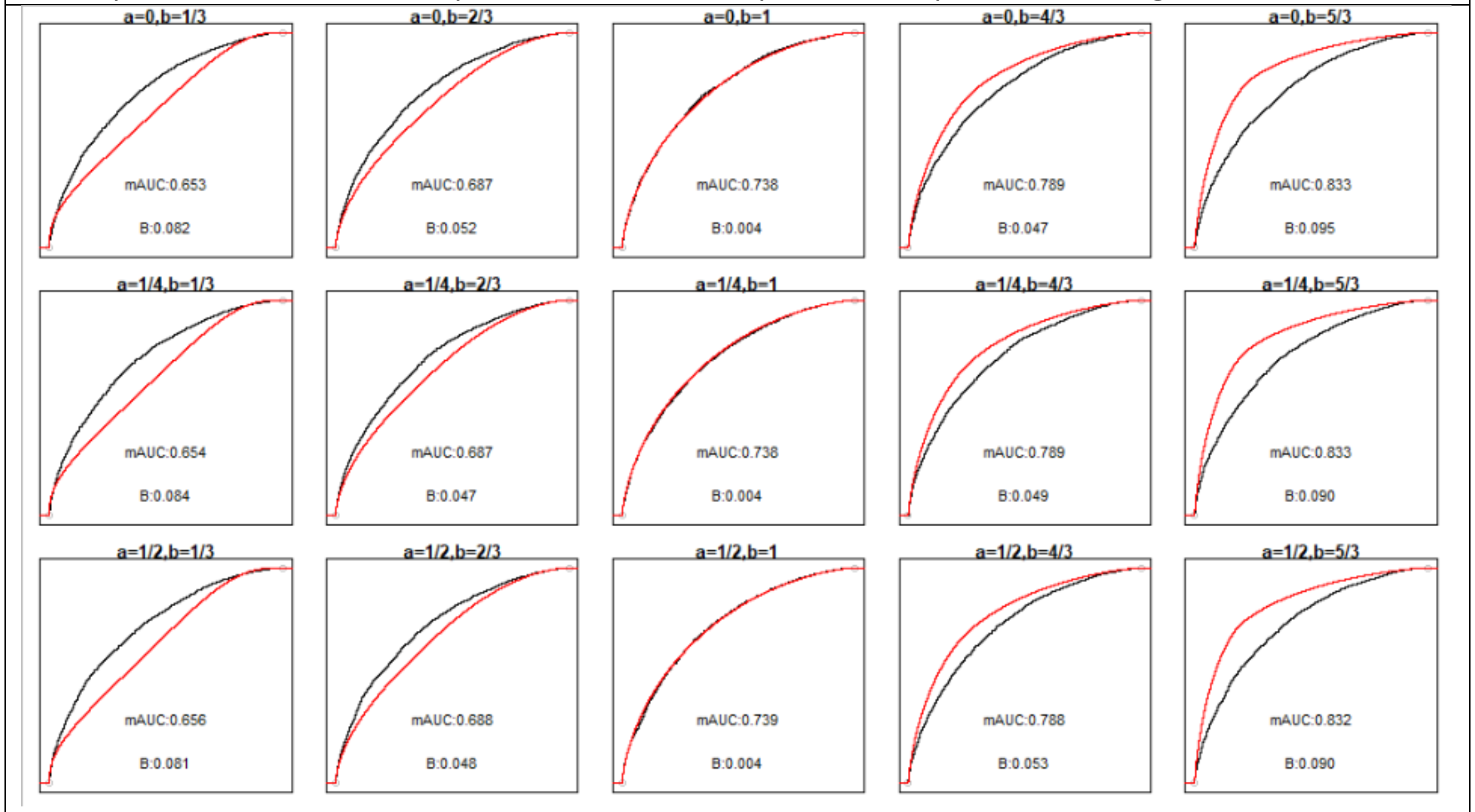

The ROC curves approximate the population-level curves as they are based on a large sample size (10,000 simulated observations). The area under the ROC curve is 0.740 in all scenarios.

**ROC**: Receiver Operating Characteristic; $\boldsymbol{B}$: ROC equality statistic; mAUC: area under the model-based ROC curve

***Figure 4:*** Probability of rejecting the null hypothesis for the mean calibration (first white bar), ROC equality (second white bar), and unified (orange bar) test statistics, along with the Hosmer-Lemeshow test (purple bar). The panels positionally correspond to the calibration plots and simulation parameters presented in Figure 2.

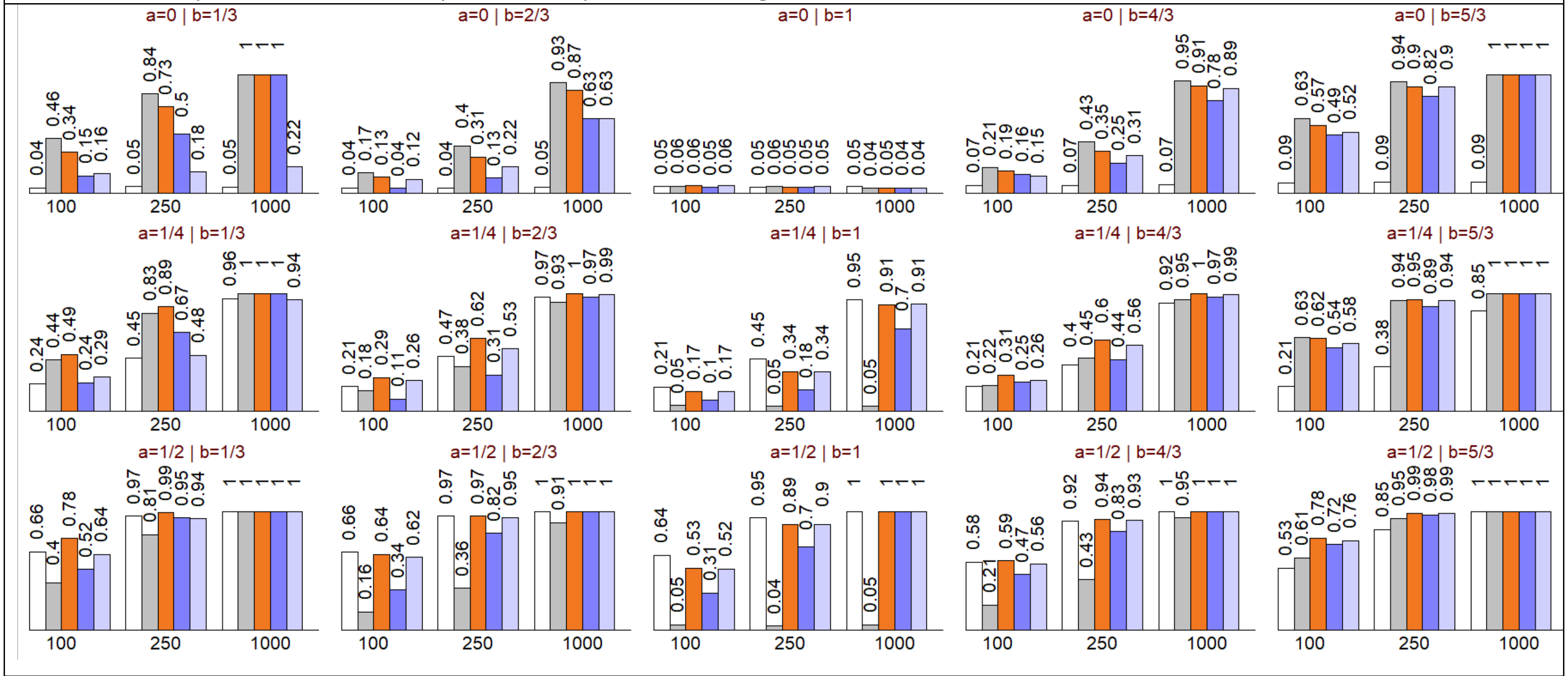

**ROC**: Receiver Operating Characteristic

***Figure 5:*** The empirical ROC curves from the MACRO development (blue) and STATCOPE validation (black) samples, the mROC curve from the STATCOPE validation sample (red) (left panels) and the calibration plot (right panels).

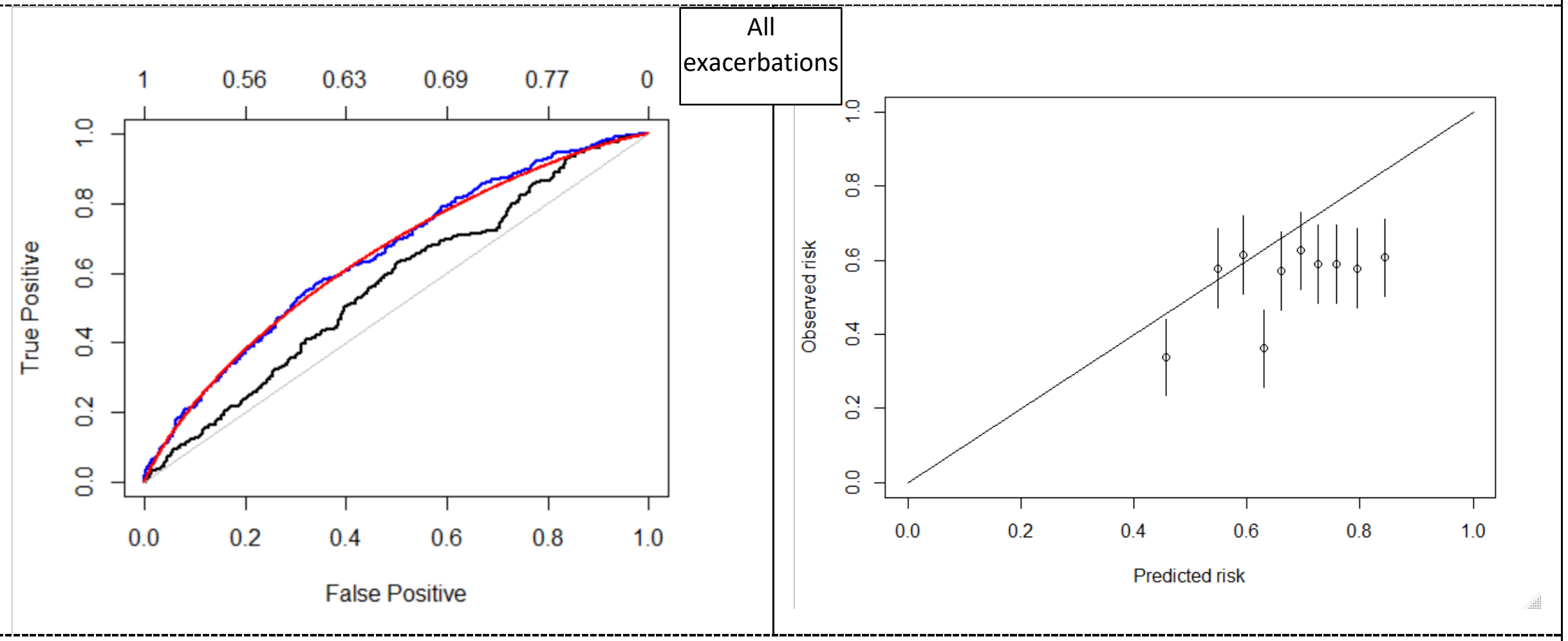

$AUC_{dev} = 0.65; AUC_{val} = 0.56;\ \ mAUC = 0.65; A_n = 0.125\ (observed < predicted);\ B_n = 0.082$

$p_{A_n} < 0.001;\ \ p_{B_n} < 0.001;\ \ p_{U_n} < 0.001$

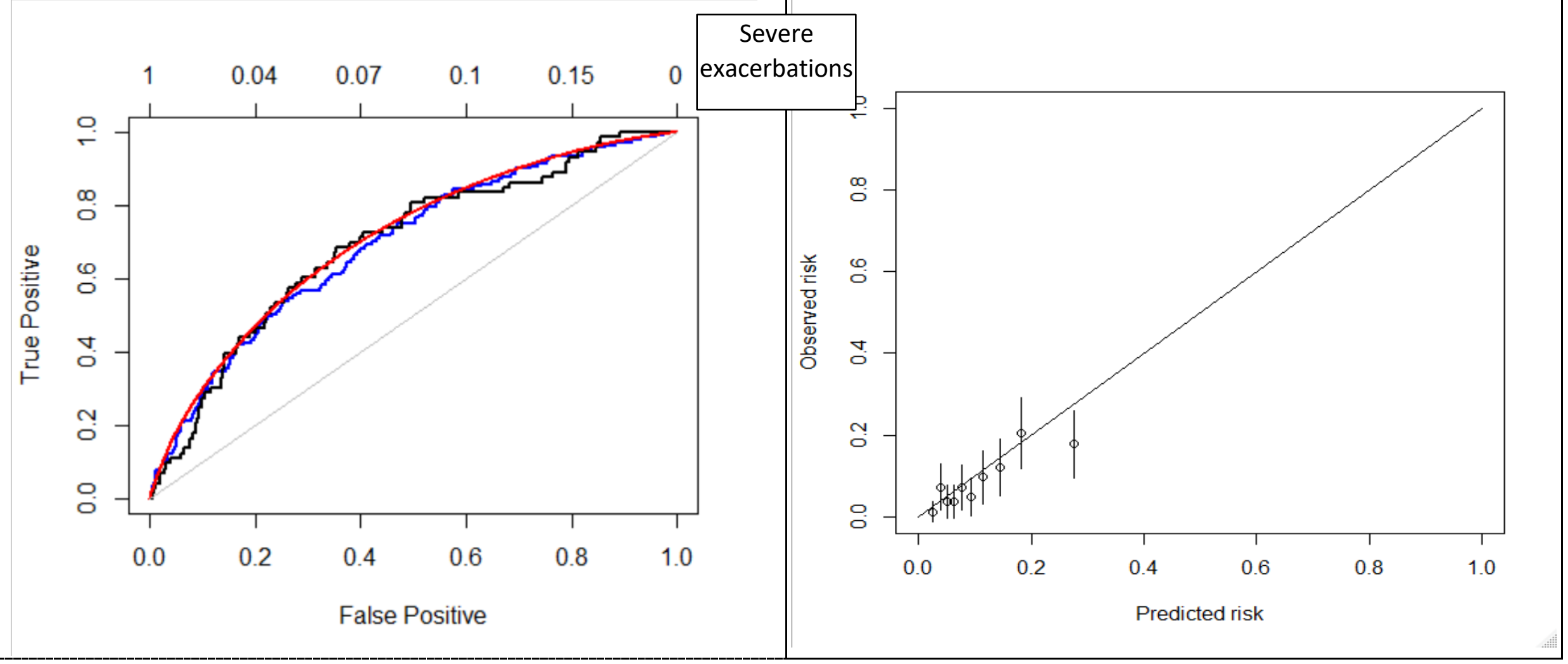

| |
|---|
| $AUC_{dev} = 0.69; AUC_{val} = 0.69;\ mAUC = 0.707;\ A_n = 0.019\ (observed < predicted); B_n = 0.022$<br>$p_{A_n} = 0.070;\ p_{B_n} = 0.74;\ p_{U_n} = 0.20$ |
| The second x-axis (top) indicates the cutoff points on predicted risk corresponding with the empirical ROC for the validation sample.<br><br>$\boldsymbol{AUC_{dev}}$: area under the curve (c-statistic) in the development sample; $\boldsymbol{AUC_{val}}$ : area under the curve (c-statistic) in the validation sample; $\boldsymbol{mAUC}$: area under the model-based ROC curve; $\boldsymbol{A_n}$: Mean calibration statistic; $\boldsymbol{B_n}$: ROC equality statistic; $\boldsymbol{p_{A_n}}$: P value of the mean calibration test; $\boldsymbol{p_{B_n}}$: P value for the ROC equality test; $\boldsymbol{p_{U_n}}$: P value of the unified test, |

# Supplementary Material for

## Model-based ROC (mROC) curve: examining the effect of case-mix and model calibration on the ROC plot

Mohsen Sadatsafavi; Paramita Saha-Chaudhuri; John Petkau

**Table of content**

**Section 1:** Proof of the equivalence of area under mROC and model-based c-statistic

Proposition: Let $f(.)$ be the probability density function of the random variable X. Let $G(.)$ be the cumulative distribution function of an independent random variable Y. If both X and Y take values in [0,1], then

$$P(X > Y) = \int_0^1 P(X > Y|X = z)dP(X \leq z) = \int_0^1 P(Y < z)f(z)dz = \int_0^1 f(z)G(z)dz$$

Proof of the main lemma: As in the main text, let Y* be a model-based response, and define $\bar{F}_0$ and $\bar{F}_1$ as the CDFs underlying mROC (the CDFs of predicted risks among individuals with Y* of 0 and 1, respectively). Let $\pi_0^*$ and $\pi_1^*$ be random draws from these two distributions . The model-based c-statistic (mbc) is the probability that among two individuals with discordant model-based responses, the one with $Y^* = 1$ has a higher predicted risk than that with $Y^* = 0$. That is, $mbc = P(\pi_1^* > \pi_0^*)$.

As for the mROC, we have:

$$mROC(t) = 1 - \bar{F}_1\big(\bar{F}_0^{-1}(1 - t)\big),$$

where 0 ≤ $t$ ≤ 1 is the false positive probability, so

$$mAUC = \int_0^1 \big\{1 - \bar{F}_1\big(\bar{F}_0^{-1}(1 - t)\big)\big\}dt.$$

The change of variable $x = \bar{F}_0^{-1}(1 - t)$ leads to

$$mAUC = \int_0^1 \{1 - \bar{F}_1(x)\}\bar{f}_0(x)dx = 1 - P(\pi_0^* > \pi_1^*) = P(\pi_1^* > \pi_0^*) = mbc.$$

**Section 2:** Proof of the convergence of mROC and empirical ROC curves under model calibration

**Lemma:** For a moderately calibrated risk prediction model, the empirical and model-based ROC curves asymptotically converge.

**Proof:** Let $\boldsymbol{X}$ be the vector of covariates (predictors), with $\mathbf{X}_i$ referring to the realization of this vector for the i^th individual. A pre-specified risk prediction model $\pi^*(\mathbf{X})$ yields predicted risks $\pi^* \equiv \pi^*(\mathbf{X}_i)$. When sampling from a population, the mapping from $\mathbf{X}_i$ to $\pi_i^*$ is known, but $\pi_i^*$ for the ith individual is random as $\mathbf{X}_i$ is randomly selected. For any value of the predicted risk $\pi^*$, there is a unique 'calibrated risk' $\pi$ given by the true risk of the outcome among all individuals with that predicted risk: $\pi \equiv \pi(\pi^*) = P(Y = 1|\pi^*(\mathbf{X}) = \pi^*)$. A model is moderately calibrated when $\forall z, \pi(z) = z$.

We first consider the behavior of $F_{1n}(t)$. For each fixed value of $t$, $F_{1n}(t)$ is the average of $I(\pi_i^* \leq t)$ among individuals with $Y_i = 1$. Hence, provided $P(Y = 1) > 0$, dividing both the numerator and denominator of the expression for $F_{1n}(t)$ in the main text by $n$ and applying the Weak Law of Large Numbers (in what follows, an arrow denotes convergence in probability as the sample size *n* approaches infinity), yields:

$$F_{1n}(t) \to \frac{E[I(\pi^* \leq t).Y]}{E(Y)} = \frac{P(\pi^* \leq t,\ Y = 1)}{P(Y = 1)} = P(\pi^* \leq t\,|Y = 1) = F_1(t).$$

Bayes' rule allows this limit to be re-expressed as

$$P(\pi^* \leq t \mid Y = 1) = \frac{P(Y = 1 \mid \pi^* \leq t).P(\pi^* \leq t)}{P(Y = 1)}.$$

Proceeding similarly for $\bar{F}_{1n}(t)$ leads to

$$\bar{F}_{1n}(t) \to \frac{E[I(\pi^* \leq t).\pi^*]}{E[\pi^*]} = \frac{P(\pi^* \leq t, Y^* = 1)}{P(Y^* = 1)} = P(\pi^* \leq t \mid Y^* = 1) = \bar{F}_1(t).$$

Again, applying the Bayes' rule, we have

$$P(\pi^* \leq t\,|Y^* = 1) = \frac{P(Y^* = 1 \mid \pi^* \leq t).P(\pi^* \leq t)}{P(Y^* = 1)}.$$

For a moderately calibrated risk prediction model where $\pi(\pi^*) = \pi^*$ , it follows immediately that $P(Y = 1) = E(\pi) = E(\pi^*) = P(Y^* = 1)$. To prove $F_{1n}(t) - \bar{F}_{1n}(t) \to 0$ we therefore only need to show that $P(Y = 1 \mid \pi^* \leq t) - P(Y^* = 1 \mid \pi^* \leq t) = 0$. But we have $P(Y = 1 \mid \pi^* \leq t) - P(Y^* = 1 \mid \pi^* \leq t) \propto \int_0^t \{P(Y = 1|\pi^* = z) - P(Y^* = 1|\pi^* = z)\}. dP(\pi^* \leq z) = 0$, by the definition of moderate calibration.

Similar arguments apply for $F_{0n}(t)$ and $\bar{F}_{0n}(t)$, thereby establishing the desired result.

**Section 3:** A stylized example demonstrating the connection between mROC and model calibration

Consider the simple situation when the true risk, represented by $p$, has a standard uniform distribution in the population:

$$p \sim uniform(0,1),$$
$$Y \sim Bernoulli(p).$$

We consider three scenarios: the 'correct specification' scenario, when the prediction model correctly estimates the true risk ($\pi^* = p$) and thus is calibrated, and two alternative scenarios of overestimation ($\pi^* = \sqrt{p}$) and underestimation ( $\pi^* = p^2$) of the true risks. For these three scenarios, the analytical forms of the population-based CDFs $F_1(t)$, $F_0(t)$ , $\bar{F}_1(t)$, and $\bar{F}_0(t)$ are provided in ***Table S1***.

***Table S1***: Population-based forms of the cumulative distribution functions underlying the empirical and model-based ROC curves for the simple uniform risk situation

| **CDF** | **Scenarios** | | |
|---|---|---|---|
| | Correct specification | Overestimated risk | Underestimated risk |
| | $\pi^* = p$ | $\pi^* = \sqrt{p}$ | $\pi^* = p^2$ |
| $F_1(t)$ | $t^2$ | $t^4$ | $t$ |
| $F_0(t)$ | $2t - t^2$ | $2\,t^2 - t^4$ | $2\sqrt{t} - t$ |
| $\bar{F}_1(t)$ | $t^2$ | $t^3$ | $t^{3/2}$ |
| $\bar{F}_0(t)$ | $2t - t^2$ | $3\,t^2 - 2\,t^3$ | $(3\sqrt{t} - t^{\frac{3}{2}})\,/\,2$ |

For all three scenarios, given that the predicted risks are monotonically transformed versions of the true risk, the population-based ROCs are the same: $ROC(t) = 2\sqrt{t} - t$. However, $mROC(t) = ROC(t)$ only for the correct specification scenario. For the two alternative scenarios, closed-form expressions for $mROC(t)$ are not available, but the single root for $\bar{F}_0{}^{-1}(1-t)$ can be found numerically to evaluate $mROC(t)$. Results are provided in ***Figure S1***, where the ROC, mROC, and corresponding population-based calibration plots are provided for comparison. The latter have closed-form expressions in this simple situation, but in general calibration plots cannot be drawn without grouping or smoothing the data. On the other hand, the mROC curve can be evaluated from a sample without the requirement for any such arbitrary specifications.

***Figure S1:*** ROC and mROCs for the simple uniform risk situation (left) and the corresponding calibration plots (right). Black: fully calibrated model; red: over-estimated risk; blue: underestimated risk. For the left panel, the ROC curves coincide (black line) for the three scenarios considered, but the mROC curves are distinct. For the right panel, the calibration curves for the three scenarios are all distinct.

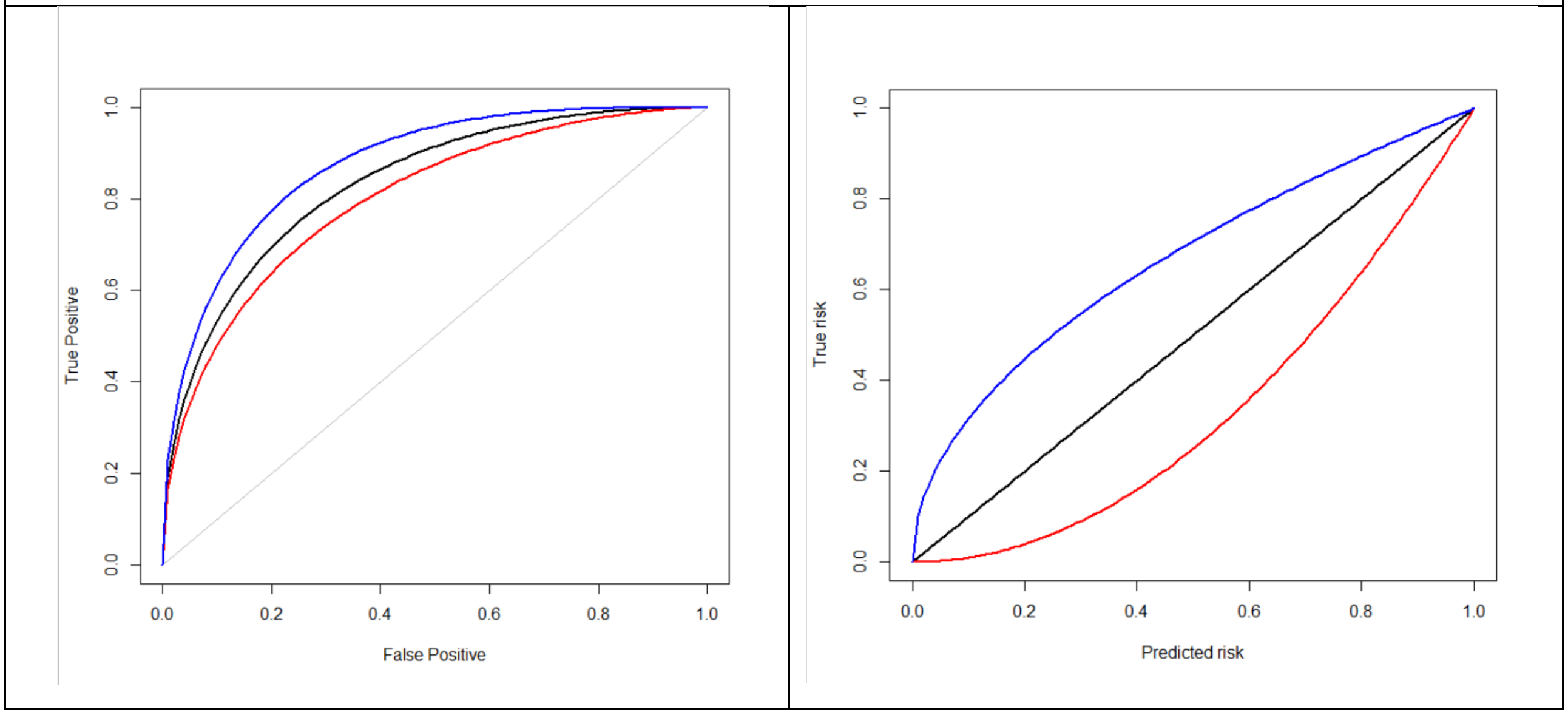

**Section 4:** Sufficient conditions for moderate calibration

**Lemma:** If the expected value of the predicted and true risks are the same in the population, then pointwise equality of population ROC and mROC curves implies the model is at least moderately calibrated.

**Proof:** Let $\pi^* = \pi^*(\mathbf{X})$ represent the predicted risk, and $G^*(\cdot)$ its CDF. Let $\pi(\cdot)$ be the true calibration function, representing the mapping from $\pi^*$ to the actual risk: $\pi(z) = P(Y = 1|\pi^* = z)$. A model being at least moderately calibrated means $\pi(z) = z$ almost everywhere (a.e.) on the support of $G^*(\cdot)$.

Given that the result to be established is concerned with population quantities, in place of the CDFs $F_{1n}(t)$, $F_{0n}(t)$, $\bar{F}_{1n}(t)$, and $\bar{F}_{0n}(t)$ that underlie the empirical ROC and mROC curves, we use the limiting versions of these CDFs.

For the ROC curve, we can express the underlying CDFs as

$$F_1(t) = P(\pi^* \leq t|Y = 1) = \frac{P(\pi^* \leq t, Y = 1)}{P(Y = 1)} = \frac{E[I(\pi^* \leq t).\pi(\pi^*)]}{E[\pi(\pi^*)]} = \frac{\int_0^t \pi(u).dG^*(u)}{\int_0^1 \pi(u).dG^*(u)},$$

and similarly,

$$F_0(t) = \frac{\int_0^t \big(1 - \pi(u)\big).dG^*(u)}{1 - \int_0^1 \pi(u).dG^*(u)}.$$

For the mROC curve, similar derivations result in

$$\bar{F}_1(t) = \frac{\int_0^t u.dG^*(u)}{\int_0^1 u.dG^*(u)},$$

and

$$\bar{F}_0(t) = \frac{\int_0^t (1-u).dG^*(u)}{1-\int_0^1 u.dG^*(u)}.$$

For the sake of simplicity and to avoid technicalities around the behavior of the quantile function for discrete distributions, the proof presented here is for the common case where $G^*(\cdot)$ is a strictly increasing function without jumps (equivalently, it has a corresponding probability density function having no intervals with zero density). This is the case, for example, for typical logistic regression models when there is at least one continuous predictor with unrestricted range. Given this condition, $\bar{F}_1(t)$ and $\bar{F}_0(t)$ are strictly increasing (without jumps) on [0,1] and, with the additional technical condition that $0 < \pi(z) < 1$ (the true risk is not strictly 0 or 1 at any level of predicted risk), so too are $F_1(t)$ and $F_0(t)$.

With these expressions, we can re-express the result to be established as

$$\begin{cases} \mathit{Condition}\ 1:\ \int_0^1 u.dG^*(u) = \int_0^1 \pi(u).dG^*(u) \\ \mathit{Condition}\ 2: \forall t\ \bar{F}_1\left(\bar{F}_0^{-1}(1-t)\right) = F_1\left(F_0^{-1}(1-t)\right) \end{cases} \Longrightarrow \pi(z) = z\ a.e.$$

Let $a = a(t) = \bar{F}_0^{-1}(1-t)$ and $= b(t) = F_0^{-1}(1-t)$; it follows that $\bar{F}_0(a) = F_0(b)$. Then *Condition 2*, and the strictly increasing nature of the CDFs, imply:

$$\bar{F}_0(a) = F_0(b) \Leftrightarrow \bar{F}_1(a) = F_1(b).$$

The expressions above for these CDFs yield the equivalent statement (after making use of *Condition 1*) that, for each fixed $t$:

$$\int_0^a (1-u).dG^*(u) = \int_0^b [1-\pi(u)].dG^*(u) \Leftrightarrow \int_0^a u.dG^*(u) = \int_0^b \pi(u).dG^*(u),$$

or equivalently,

$$\int_0^a u.dG^*(u) = \int_0^b \pi(u).dG^*(u) \Leftrightarrow G^*(a) = G^*(b).$$

Let $G^{*-1}(.)$ be the quantile function of $G^*(.)$. Setting $x = x(t) = G^*(a) = G^*(b)$, the previous statement can be written as:

$$\forall x \int_0^{G^{*-1}(x)} u.\,dG^*(u) = \int_0^{G^{*-1}(x)} \pi(u).\,dG^*(u).$$

With a change of variable $y = G^*(u)$, this becomes:

$$\forall x \int_0^{x} G^{*-1}(y).\,dy = \int_0^{x} \pi\left(G^{*-1}(y)\right).\,dy,$$

implying that $\pi(z) = z$ almost everywhere on the support of $G^*(\cdot)$, the probability distribution of the predicted risks.

**Section 5:** Calculating a unified P value for the assessment of model calibration

1. Calculate $A_n$ and $B_n$ from the vectors of $\boldsymbol{\pi}^*$ and $\mathbf{Y}$. These are the point estimates of the test statistics.
2. For i=1 to N (number of simulations):
   2.1. Generate a random response vector $\mathbf{Y}_i^*$ from the predicted risks $\boldsymbol{\pi}^*$.
   2.2. Calculate $A_{0i}$ and $B_{0i}$ from $\boldsymbol{\pi}^*$and $\mathbf{Y}_i^*$and store their values.
3. Based on the $A_{0i}s$ and $B_{0i}s$, construct the empirical CDFs $eCDF_{A_n}(.)$ and $eCDF_{B_n}(.)$.
4. Calculate $p_{A_n} = 1 - eCDF_{A_n}(A_n)$, $p_{B_n} = 1 - eCDF_{B_n}(B_n)$, and $U_n = -2.\left[\log(p_{A_n}) + \log(p_{B_n})\right]$.
5. For each simulated vector $\mathbf{Y}_i^*$, use the same empirical CDFs to calculate simulated p-values $p_{A_i}$, $p_{B_i}$, and test statistic $U_{n_i}$. For these N values of $U_{n_i}$, calculate $c = \frac{var(U_n)}{2.average(U_n)}$ and $k = \frac{2.average(U_n)^2}{var(U_n)}$.
6. The unified P value is evaluated as $p_{U_n} = 1 - F\left(\frac{U_n}{c}; k\right)$, where $F(.;k)$ is the CDF of the chi-square distribution with $k$ degrees of freedom.

**Section 6:** Results of the first set of simulations

The simulation setup was similar to that of the main text. We generated a single predictor $X \sim Normal(0,1)$, and modeled the true risk as $p = 1/(1 + exp(-X))$, resulting in the population average response probability of 0.5. We then evaluated the performance of the test in a simulated independent sample of $n$ observations when the predicted risks suffer from various degrees of mis-calibration. This was modeled by applying a logit-linear transformation of the true risks to generate the predicted risks: $logit(\pi^*) = a + b.X$. We simulated response values and predicted risks under a fully factorial design with values $a = \{-0.25, -0.125, 0, 0.125, 0.25\}, b = \{0.5, 0.75, 1, 1.5, 2\}$, creating 25 simulation scenarios each for $n = \{100, 250, 1000\}$.

In this particular setup, if $Y$ is the observed binary response, a likelihood ratio test for $\beta_0 = 0$ and $\beta_1 = 1$ in the logistic model $logit(P(Y = 1)) = \beta_0 + \beta_1.logit(\pi^*)$, can be used to test for moderate calibration. This model is equivalent to a typical logistic regression model where the logit-transformed predicted probabilities are considered a covariate. In general, the likelihood ratio test is a test for 'weak calibration' in the hierarchical definition of model calibration proposed by Van Calster et al, with weak calibration achieved if $\beta_0 = 0$ and $\beta_1 = 1$(1). However, in this setup, it is a valid test for moderate calibration because the link function (logit) is known and the associations are known to be linear on the logit scale (therefore weak and moderate calibration are equivalent in this setting). As such, and according to the Neyman-Pearson lemma, the likelihood ratio test is the most powerful test for moderate calibration in this setup, providing a yardstick to evaluate the performance of the proposed test.

The relationships between the predicted and true risks are depicted in ***Figure S2.*** The ROC and mROC curves are presented in ***Figure S3***. Results of the simulation studies, in terms of the proportion of times the null hypotheses were rejected, are provided in ***Figure S4.*** As these results demonstrate, the unified test performs very similarly to the likelihood ratio test.

***Figure S2:*** Relationship between predicted (X axis) and true (Y axis) risks.

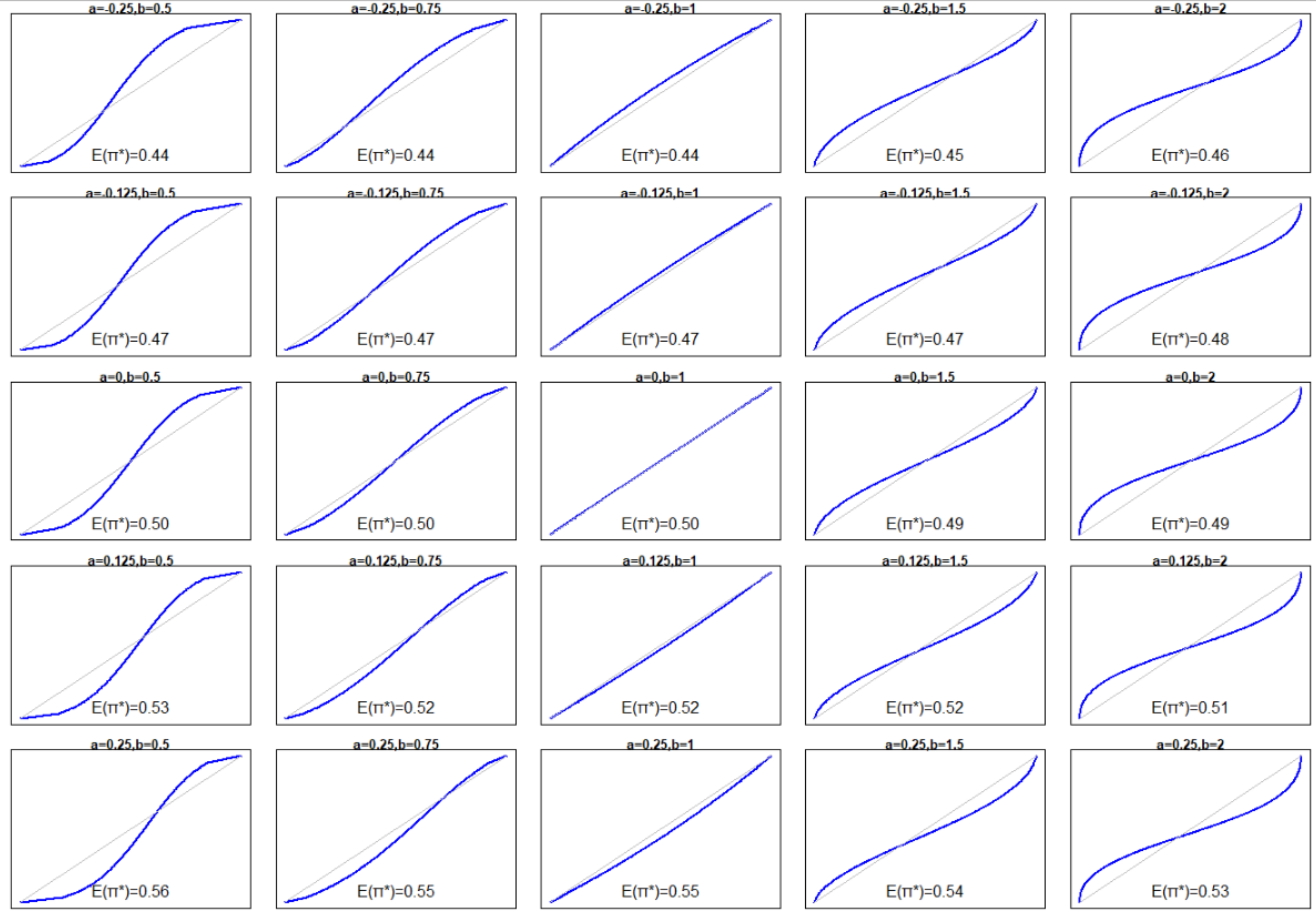

***Figure S3:*** ROC (black) and mROC (red) curves for the simulation scenarios. The panels positionally correspond to the calibration plots and simulation parameters presented in Figure S2.

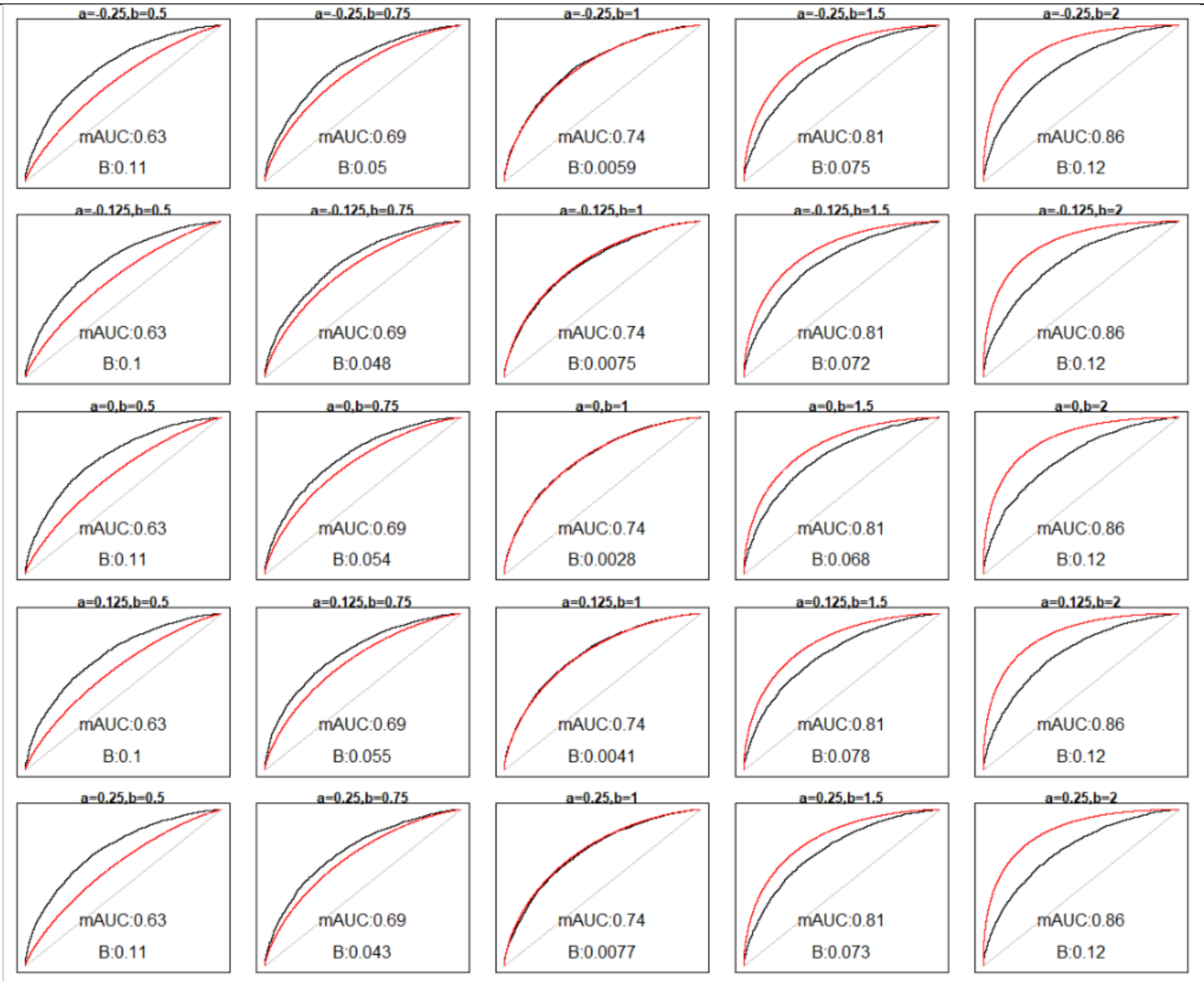

***Figure S4:*** Probability of rejecting the null hypothesis at 0.05 level for the mean calibration (pink bars), ROC equality (orange bars), unified (purple bars), and likelihood ratio (white bars) tests. The panels positionally correspond to the calibration plots and simulation parameters presented in Figure S2.

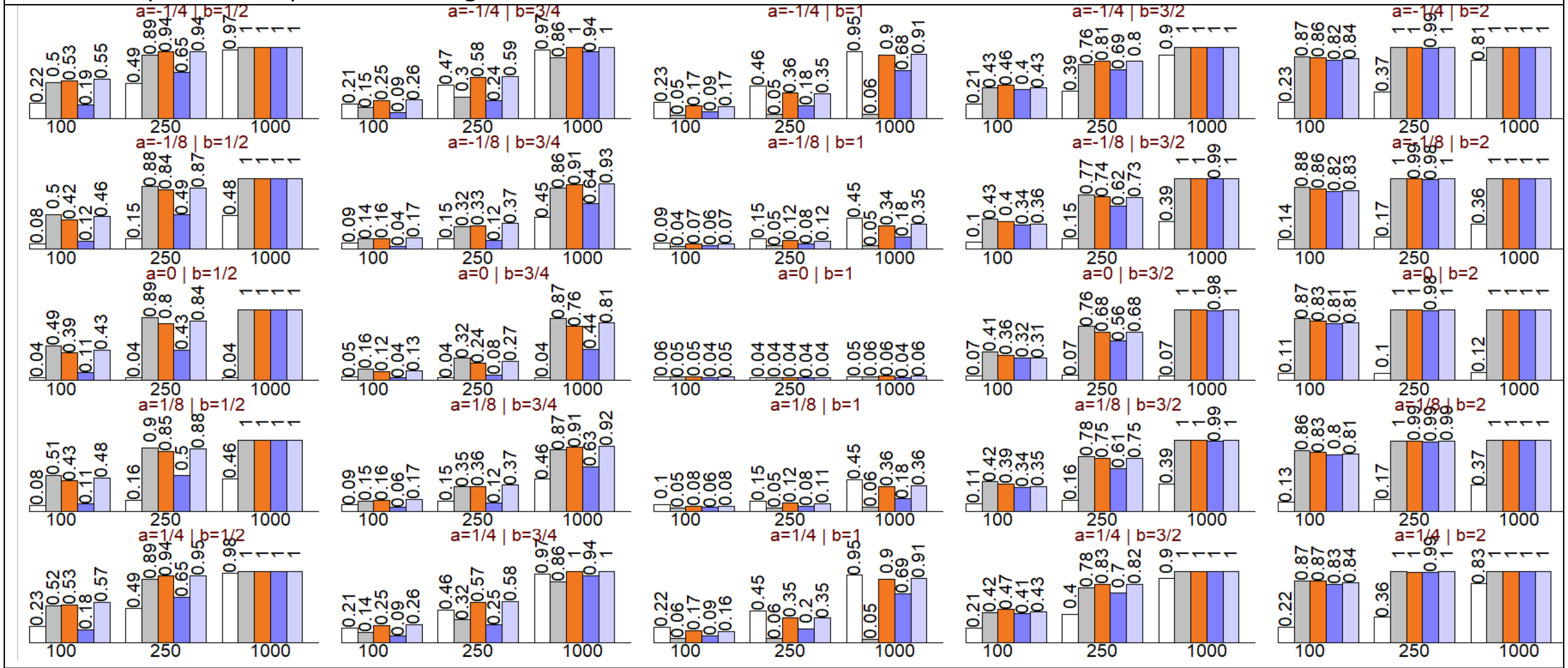